\documentclass[12pt]{iopart}

\usepackage{graphicx}
\usepackage{dcolumn}
\usepackage{bm}

\def\cpd{\rm{kg^{-1}keV^{-1}day^{-1}}}
\def\nuebar{\rm{\bar{\nu}_e}}

\def\ge73m{\rm{^{73}Ge^{*} ( 1/2 ^{-} ) }}
\def\s2tw{\rm{ sin ^2 \theta _W }}
\def\phigeb{\rm{\Phi_{Ge}^{-}}}
\def\phigea{\rm{\Phi_{Ge}^{+}}}
\def\phinaib{\rm{\Phi_{NaI}^{-}}}
\def\phinaia{\rm{\Phi_{NaI}^{+}}}

\begin{document}

\hfill AS-TEXONO/08-02\\
\hspace*{1cm} \hfill \today

\title[]{Production and Decay 
of the $\ge73m$ Metastable State in 
a Low-Background Germanium Detector}

\author{
H.Y.~Liao$^{1,2}$,
H.M.~Chang$^{1,2}$,  
M.H.~Chou$^{1}$,  
M.~Deniz$^{1,3}$, 
H.X.~Huang$^{1,4}$,  
F.S.~Lee$^{1}$,  
H.B.~Li$^{1}$,   
J.~Li$^{5,6}$,  
C.W.~Lin$^{1}$, 
F.K.~Lin$^{1}$,
S.K.~Lin$^{1}$,  
S.T.~Lin$^{1}$,  
V.~Singh$^{1}$,  
H.T.~Wong$^{1, \ast}$,  
S.C.~Wu$^{1}$
}

\vspace*{0.5cm}

\address{
{\normalsize TEXONO Collaboration:}\\
$^1$ {Institute of Physics, Academia Sinica, Taipei 115, Taiwan.}\\
$^2$ {Department of Physics, National Taiwan University,
Taipei 106, Taiwan.}\\
$^3$ {Department of Physics,
Middle East Technical University, Ankara 06531, Turkey.}\\
$^4$ {Department of Nuclear Physics,
Institute of Atomic Energy, Beijing 102413, China.}\\
$^5$ {Institute of High Energy Physics,
Chinese Academy of Science, Beijing 100039, China.}\\
$^6$ {Department of Engineering Physics, Tsing Hua University,
Beijing 100084, China.}\\[2ex]
$^{\ast}$ {Corresponding Author$-$ Email: htwong@phys.sinica.edu.tw;
Tel:+886-2-2789-9682; FAX:+886-2-2788-9828.}\\
}

\begin{abstract}

The $\ge73m$ metastable state decays with a very characteristic
signature which allows it to be tagged event-by-event.
Studies were performed using data taken with a
high-purity germanium detector in a low-background
laboratory near a nuclear power reactor core
where the $\nuebar$-flux  was
$\rm{6.4 \times 10^{12} ~ cm^{-2} s^{-1}}$.
The measured average and equilibrium 
production rates of $\ge73m$
were $\rm{( 8.7 \pm 0.4 )}$ and
$\rm{( 6.7 \pm 0.3 ) ~ kg^{-1} day^{-1}}$, respectively. 
The production channels were studied and identified.
By studying the difference in the production
of $\ge73m$ between the reactor ON and OFF spectra,
limiting sensitivities 
in the range of $\rm{\sim 10^{-42} - 10^{-43} ~ cm^2}$ 
for the cross sections 
of neutrino-induced nuclear transitions 
were derived.
The dominant background are due to $\beta ^-$ decays
of cosmic-ray induced $^{73}$Ga. 
The prospects of enhancing the sensitivities 
at underground locations are discussed.

\end{abstract}


\maketitle

\section{Introduction}

A research program
on low energy neutrino physics~\cite{pdgnuphys}
is being pursued by the TEXONO
Collaboration at the Kuo-Sheng (KS) Reactor Laboratory
in Taiwan~\cite{ksprogram}.
A search for the neutrino magnetic moment~\cite{magmommpla}
was performed  with an ultra-low-background
high-purity germanium detector (HPGe),
from which a limit of
$\mu_{\nu} ( \nuebar ) < 7.4 \times 10^{-11} ~ \mu_{\rm B}$
at 90\% confidence level (CL) was
derived~\cite{texonomagmom}.
During the course of this study, the unique signature
of the decay of the $\ge73m$ metastable state
was noted.

In this report,  we discuss 
these decay signatures 
in Section~\ref{sect::expt}
and present studies
of their production channels
in Section~\ref{sect::prod}.
These investigations are of relevance to
the understanding of backgrounds in this
and other germanium-based low-background experiments,
such as those on 
double-beta decays~\cite{pdgdbd,dbdge} 
and cold dark-matter searches~\cite{pdgcdm,cdmge}.
Moreover, the data also allow the evaluations
of the experimental 
sensitivities of various
neutrino-induced nuclear transitions
in reactor neutrino experiments
under realistic conditions,
the studies of which are discussed in
Section~\ref{sect::nint}.

\section{Experimental Signatures of $\ge73m$ Decays}
\label{sect::expt}

\begin{figure}[hbt]
\begin{center}
\includegraphics[width=12cm]{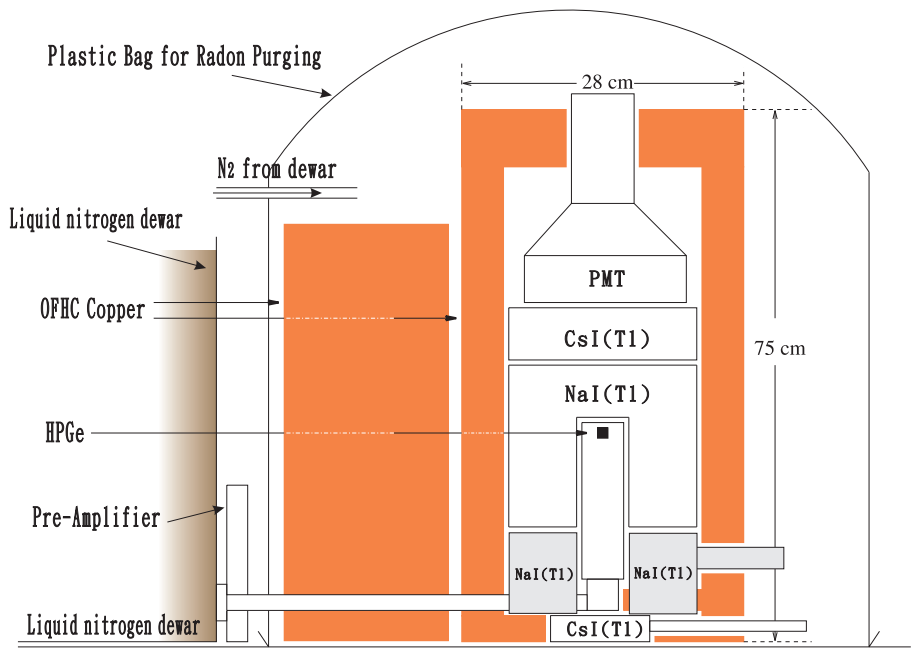}
\end{center}
\caption{
Schematic layout of the HPGe
with its anti-Compton detectors
as well as inner shieldings and
radon purge system.
}
\label{kshpge} 
\end{figure} 

\begin{figure}[hbt]
\begin{center}
\includegraphics[width=12cm]{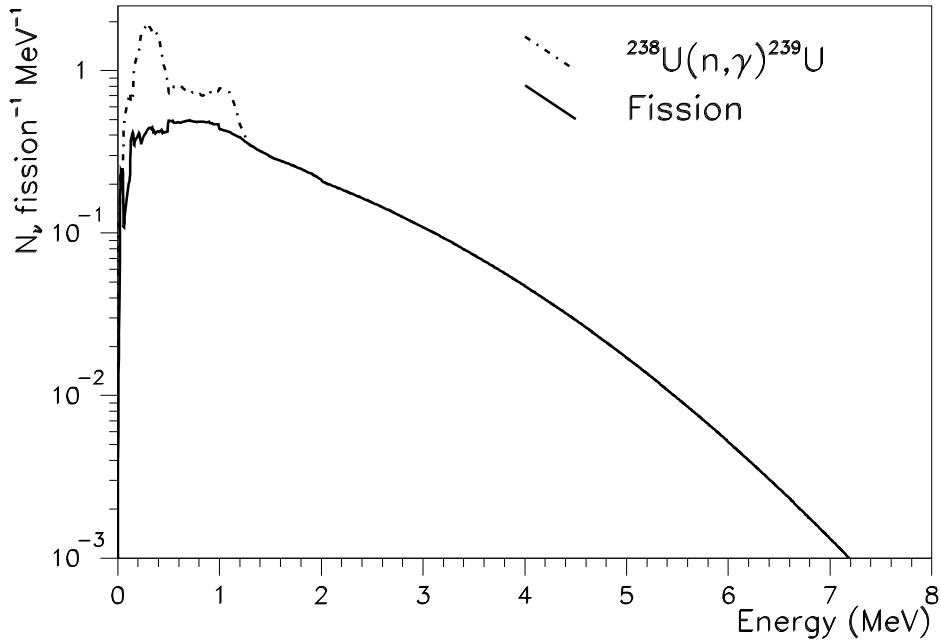}
\end{center}
\caption{
Total $\nuebar$-spectra at the power reactor,
showing the two components due to
$\beta ^-$ decays of fission fragments and
$^{239}$U.
}
\label{rnuspect}
\end{figure}

The experimental details, including detector hardware, shieldings,
electronics systems as well as studies of systematic 
uncertainties, are discussed 
in Ref.~\cite{texonomagmom}.
The schematics of the experimental set-up  is shown
in Figure~\ref{kshpge}. 
The laboratory is located at a distance of 28~m from a nuclear
reactor core with a thermal power output of 2.9~GW.
The total reactor-$\nuebar$ flux 
is about $\rm{6.4 \times 10^{12} ~ cm^{-2} s^{-1}}$.
The $\nuebar$ spectrum
is shown in Figure~\ref{rnuspect}.
The $\nuebar$'s are emitted via
$\beta ^-$ decays of 
(a) fission fragments and 
(b) $^{239}$U following neutron capture on
$^{238}$U.

\begin{figure}
\begin{center}
\includegraphics[width=14cm]{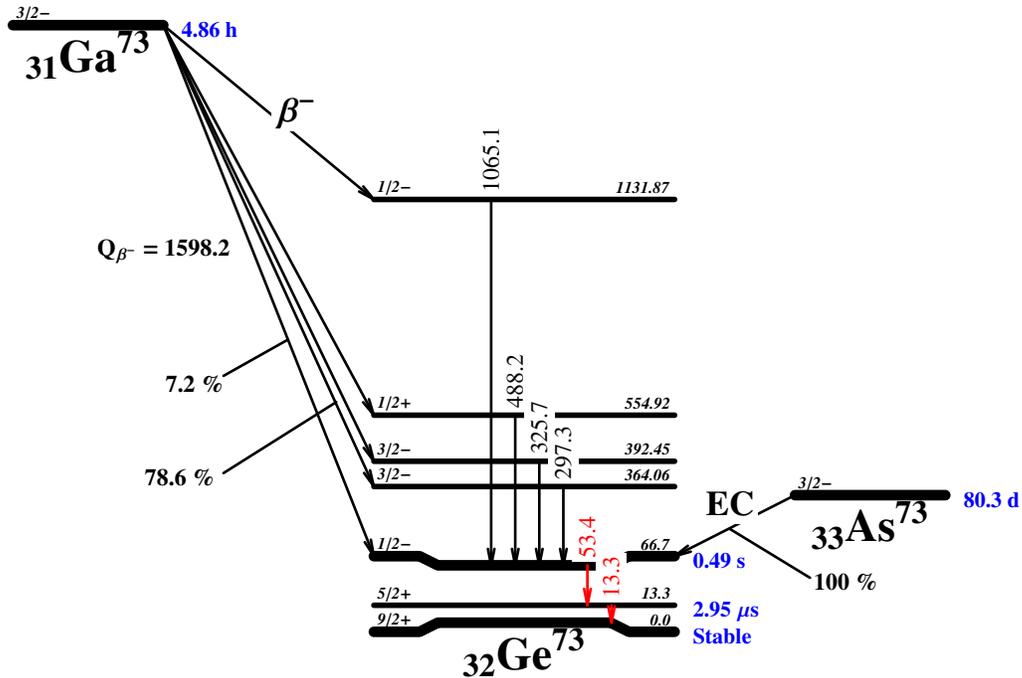}
\end{center}
\caption{
The $^{73}{\rm{Ge}}$ production 
and decay scheme~\cite{lev_scheme} 
relevant to
the discussions in this report.
}
\label{lscheme}
\end{figure}

The natural isotopic abundance of $^{73}$Ge 
in germanium is 7.73\%.
The level schemes for the isobaric states 
of $^{73}$Ga, $^{73}$Ge and $^{73}$As 
relevant to this report are 
depicted in Figure~\ref{lscheme}~\cite{lev_scheme}.
The $\ge73m$ metastable nuclei decay
via the emissions of 
53.4 keV and 13.3 keV photons 
separated by a half-life of 2.9~$\mu $s. 
This characteristic delayed coincidence
gives an experimental
signature which can be uniquely identified
in a HPGe detector.

The HPGe target was 1.06~kg in
mass and was surrounded by active anti-Compton
veto (ACV) detectors
made of NaI(Tl) and CsI(Tl) scintillating crystals.
The detector system was placed inside a 50-ton shielding
structure with an outermost layer of plastic scintillator
panels acting as cosmic-ray veto (CRV).
A physics threshold of 12~keV
and background level of $\sim 1 ~ \cpd$
comparable to those of
underground dark matter experiments
were achieved.
Besides magnetic moment searches,
this unique data set also allows the
studies of $\nu _e$~\cite{rnue} and
the searches for axions~\cite{raxion}
from the power reactor.

\begin{figure}[hbt]
\begin{center}
\includegraphics[width=12cm]{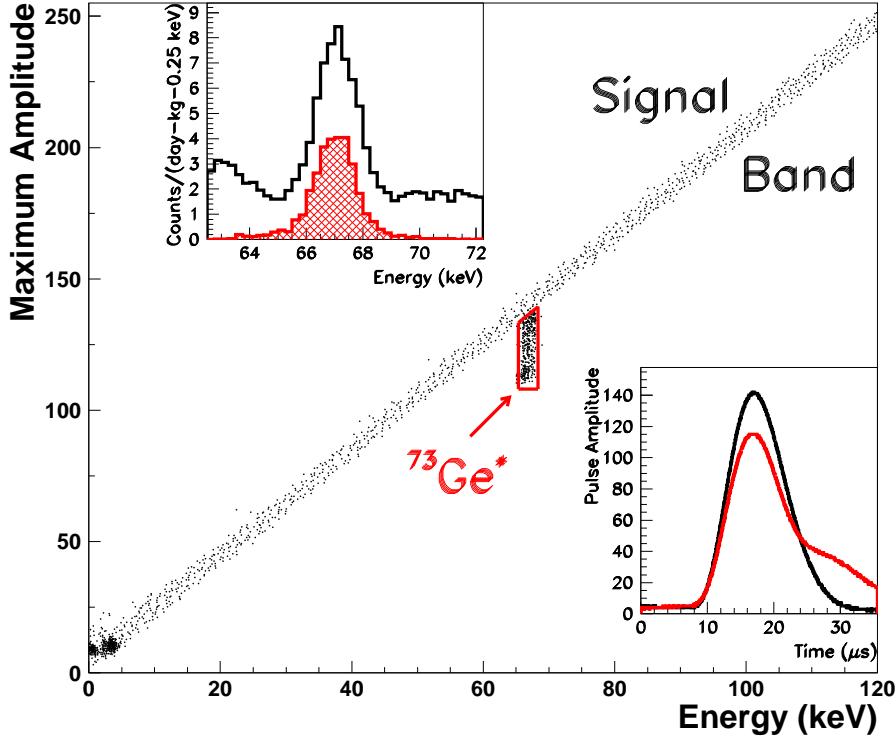}
\end{center}
\caption{
Scatter plot of 
maximum amplitude versus energy
for selected events,
showing that
decays from $\ge73m$ nuclei can be
identified perfectly.
The energy spectra before (black)
and after (red) event selection  
are shown in the left inset.
Typical pulse shapes for
normal (black) and $\ge73m$ (red)
events are
displayed in the right inset.
}
\label{ge73signature}
\end{figure}

Events uncorrelated with the CRV and ACV 
are candidates for neutrino-induced signals.
Performance and efficiencies of these selections
were thoroughly studied and documented
in Ref.~\cite{texonomagmom}.
The amplitude-versus-energy plot for the
``after-cut'' events is displayed in
Figure~\ref{ge73signature}.
The conspicuous structure at 66.7~keV
is due to the convoluted 
sum of two correlated events delayed
relative to one another,
identified unambiguously as
$\ge73m$ decays.
A typical event pair is shown
in the inset of Figure~\ref{ge73signature}.
This pulse-shape signature
is very distinct and can be tagged 
at an event-by-event basis without
contamination of background events.

Once the decays of 
$\ge73m$ were tagged,
their production channels can be studied.
The data-acquisition (DAQ) system~\cite{texonodaq} records
timing information for all events. 
This capability allows the decay sequences with half-lives
as long as 55~s to be distinctly identified~\cite{csibkg}.
The KS-P1 (180.1/52.7~days Reactor ON/OFF live time)
and 
-P3 (278.9/43.6~days Reactor ON/OFF live time)
periods, 
as defined in Ref.~\cite{texonomagmom},
were used in the analysis.
The total live times are 
459.0/96.3 days for Reactor ON/OFF,
respectively.
The selection efficiency of the $\ge73m$
events is 85\%, the
missing fraction being those with the
two $\gamma$'s emitted closer than 0.16~s 
in time to each other.
Data taken 5~s before and after these 
events were retrieved for subsequent 
detailed studies.
The production of the $\ge73m$ states
takes place mostly in the interval
of about 2~s (or 4 half-lives)
{\it before} the $\ge73m$ decays,
while events from the other intervals
represent the background control samples.
After efficiency corrections,
the production rates of 
the $\ge73m$ nuclei were measured.
The rates decreased within a DAQ period
and are characterized by two quantities.
The {\it average rates}
represent the sum of all 
$\ge73m$ decays divided by
the total DAQ live time, while 
the {\it equilibrium rates} are the
steady-state rates reached 
at the end of the DAQ periods.
The average and equilibrium rates were 
$( 8.7 \pm 0.4 ) ~ {\rm kg^{-1} day^{-1}}$
and 
$( 6.7 \pm 0.3 ) ~ {\rm kg^{-1} day^{-1}}$,
respectively.

\section{Production Channels of $\ge73m$}
\label{sect::prod}

The identified production channels for
$\ge73m$ nuclei and their relative fractions
are summarized in Table~\ref{prodsummary}.
Specific signatures of individual channels
are described in the following paragraphs.
For simplicity, 
the ``BEFORE'' and ``AFTER'' spectra
taken within the time interval (-2,0)s
and (0,2)s by detector X 
(X can be NaI or Ge)
are denoted 
by $\Phi_{X}^{-}$ and $\Phi_{X}^{+}$,
respectively.


\begin{table}
\begin{center}
\small{
\begin{tabular}{lcccccc}
\hline\hline
& \multicolumn{3}{c}{Measured Event Rate} & 
\multicolumn{3}{c}{Distribution} \\ 
& \multicolumn{3}{c}{(day$^{-1}\cdot $kg$^{-1}$)} & 
\multicolumn{3}{c}{(\%)} \\ 
\multicolumn{1}{l}{DAQ Periods} & 
P1 & P3  & Combined &
P1 & P3 & Combined 
\\ \hline \hline
\multicolumn{7}{l}{\underline{$\ge73m$ Production Rate} :} \\
~~~ DAQ End(Equilibrium)  & 8.8$\pm $0.5 & 5.2$\pm $0.4 & 6.7$\pm$0.3 & 
$-$ & $-$ & $-$ \\
~~~ Average &  9.9$\pm $0.6
& 8.1$\pm $0.5 &  8.7$\pm $0.4 &
100.0 & 100.0 & 100.0 \\ \hline \hline
\multicolumn{7}{l}{\underline{Production Channels}:} \\
\multicolumn{7}{l}{1. \underline{$^{73}$As Decays} $-$ } \\ 
~~~ DAQ Start &  
$1.8 \pm 0.2$ & 1.8$\pm $0.2 &  1.8$\pm $0.2 &
$-$ & $-$ & $-$  \\ 
~~~ DAQ End(Equilibrium)  &  
$0.3 \pm 0.1$ & 0.2$\pm $0.1 &  0.2$\pm $0.1 &
$-$ & $-$ & $-$ \\ 
~~~ Average & 1.0$\pm $0.1 & 0.7$\pm $0.1 & 0.8$\pm $0.1
& 9.8$\pm $0.9 & 9.2$\pm $0.8 & 9.5$\pm $0.6 \\
\hline
\multicolumn{7}{l}{2. \underline{$^{73}$Ga Decays} $-$ }  \\
~~~ $\beta ^{-}$ & 
2.2$\pm $0.2 & 1.9$\pm $0.2 & 2.0$\pm $0.1 &
21.7$\pm $1.6 & 23.2$\pm $1.6 & 22.5$\pm $1.1 \\ 
~~~ $\beta ^{-} \gamma$ (Full $\rm{E_{\gamma}}$) & 
0.9$\pm $0.1 & 0.7$\pm $0.1 & 0.8$\pm $0.1 &
9.1$\pm $1.0 & 9.1$\pm $0.9 & 9.1$\pm $0.6 \\ 
~~~ $\beta ^{-} \gamma$ (Partial $\rm{E_{\gamma}}$) & 
1.0$\pm $0.2 & 0.5$\pm $0.1 & 0.6$\pm $0.1 & 
9.7$\pm $1.4 & 5.6$\pm $1.0 & 7.0$\pm $0.8 \\ \hline
\multicolumn{7}{l}{3. \underline{Prompt Cosmic-Induced $\ge73m$} $-$ } \\ 
~~~ With CRV Tag & 
2.2$\pm $ 0.3 & 1.9$\pm $0.3  & 2.0$\pm $0.2 &
22.0$\pm $2.8 & 23.9$\pm $ 2.7 & 23.0$\pm $2.0 \\
~~~ HPGe only (No CRV) & 
0.0$ \pm $0.0 & 0.0$\pm $0.0  & 0.0$\pm $0.0 &
0.0$\pm $0.1 & 0.3$ \pm $0.1 & 0.1$ \pm $0.1 \\ 
~~~ HPGe+NaI (No CRV) & 
0.1$\pm $0.1 & 0.1$\pm $0.0 & 0.1$\pm $0.0 &
1.1$\pm $ 0.7 & 1.4$\pm $0.5 & 1.3$\pm $0.4 \\ \hline
\multicolumn{7}{l}
{\underline{Total Identified Production Channels for $\ge73m$} :}  \\ 
& 7.3$\pm $0.4 & 5.9$\pm $0.3 & 6.4$\pm $0.3 
& 73.4$\pm $3.8 & 72.7$\pm $3.5 & 73.0$\pm $2.6 \\  \hline \hline
\multicolumn{7}{l}
{\underline{Identified Inefficiency Factors} :}  \\
~~ DAQ Dead Time
& $-$  & $-$ & $-$
& 4.3   & 8.5  & 6.6  \\
~~ Selection Inefficiencies 
& & & & & & \\
~~~~ CRV+ACV Cuts 
& $-$ & $-$  & $-$  
& 5.0   & 8.0  & 6.6  \\
~~~~ 5~ms$<$$\Delta$t$<$=2~s
& $-$  & $-$ & $-$
& 7.1   & 7.1  & 7.1  \\
\hline \hline
\multicolumn{7}{l}{Total Identified Percentage:} \\
& $-$  & $-$ & $-$
& 89.8$\pm $3.8   & 96.3$\pm $3.5  & 93.3$\pm $2.6  \\
\hline \hline
\end{tabular}
}
\end{center}
\caption{
Summary of the 
$\ge73m$ production rates, and those
of  individual channels in the
two data-taking periods.
}
\label{prodsummary}
\end{table}


\begin{enumerate}
\item {\bf $^{73}$As :}\\
The most conspicuous difference
between the $\phigeb$ and $\phigea$ spectra,
after the CRV and ACV cuts were applied,
is the Ge X-ray
peak at 11.1~keV depicted 
in Figure~\ref{10kev_sign}.
In contrast, the structure for $\phigea$
peaks at 10.4~keV, corresponding to
the Ga X-rays following the
decays of $^{68}$Ge which were
uncorrelated with the $\ge73m$ production.
The timing distribution of the 
$\phigeb$ events associated with
the peak is shown in the inset.
The best-fit half-life is 
$(0.81 \pm 0.22) ~ {\rm s}$,
in good agreement with expectation
and demonstrating that the $\ge73m$ states
were really produced.
The time-variation of the event rates 
is depicted in 
Figure~\ref{timeplot}a.
The best-fit half-life for the 
P3 data is 
$( 62.0 \pm 23.8 ) ~ {\rm days}$,
consistent with the interpretations
of electron capture of $^{73}$As:
\begin{equation}
\rm{ ^{73}As + e^- }
 \rightarrow  \rm{ \ge73m + \nu_e + Ge~X\mbox{-}Rays } 
 ~~~ \rm{ ( ~ \tau _{\frac{1}{2}} = 80.3~days ~ ) } ~~ .
\end{equation}
This decay profile 
indicates that the production of $^{73}$As
was cosmic-ray induced.
The production rate was reduced by a factor of $\sim$9 
inside the 50-ton shielding structure
at the KS laboratory where the
overburden is about 30~meter-water-equivalence (mwe).
The steady-state equilibrium
rate of ${\rm ( 0.2 \pm 0.1 ) ~ kg^{-1} day^{-1}}$
was reached after 400~days of data taking.

\begin{figure}[hbt]
\begin{center}
\includegraphics[width=10cm]{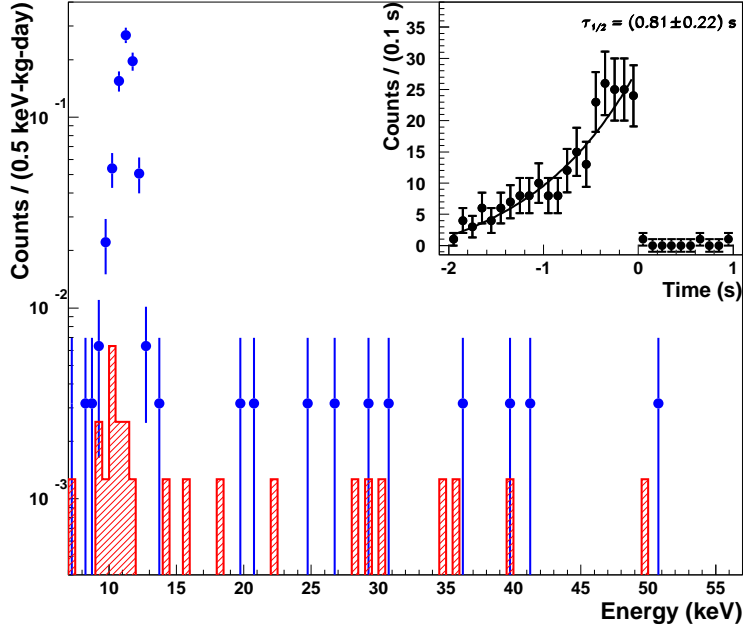}
\end{center}
\caption{
(a)
Measured HPGe $\phigeb$ (blue data points) and
$\phigea$ (red histogram) spectra 
after ACV+CRV cuts from P3 data. 
The time distribution of the Ge X-ray
events in $\phigeb$ is shown in the inset,
verifying that $\ge73m$ nuclei were produced.
}
\label{10kev_sign}
\end{figure}


\begin{figure}[hbt]
\begin{center}
{\bf (a)}\\
\includegraphics[width=10cm]{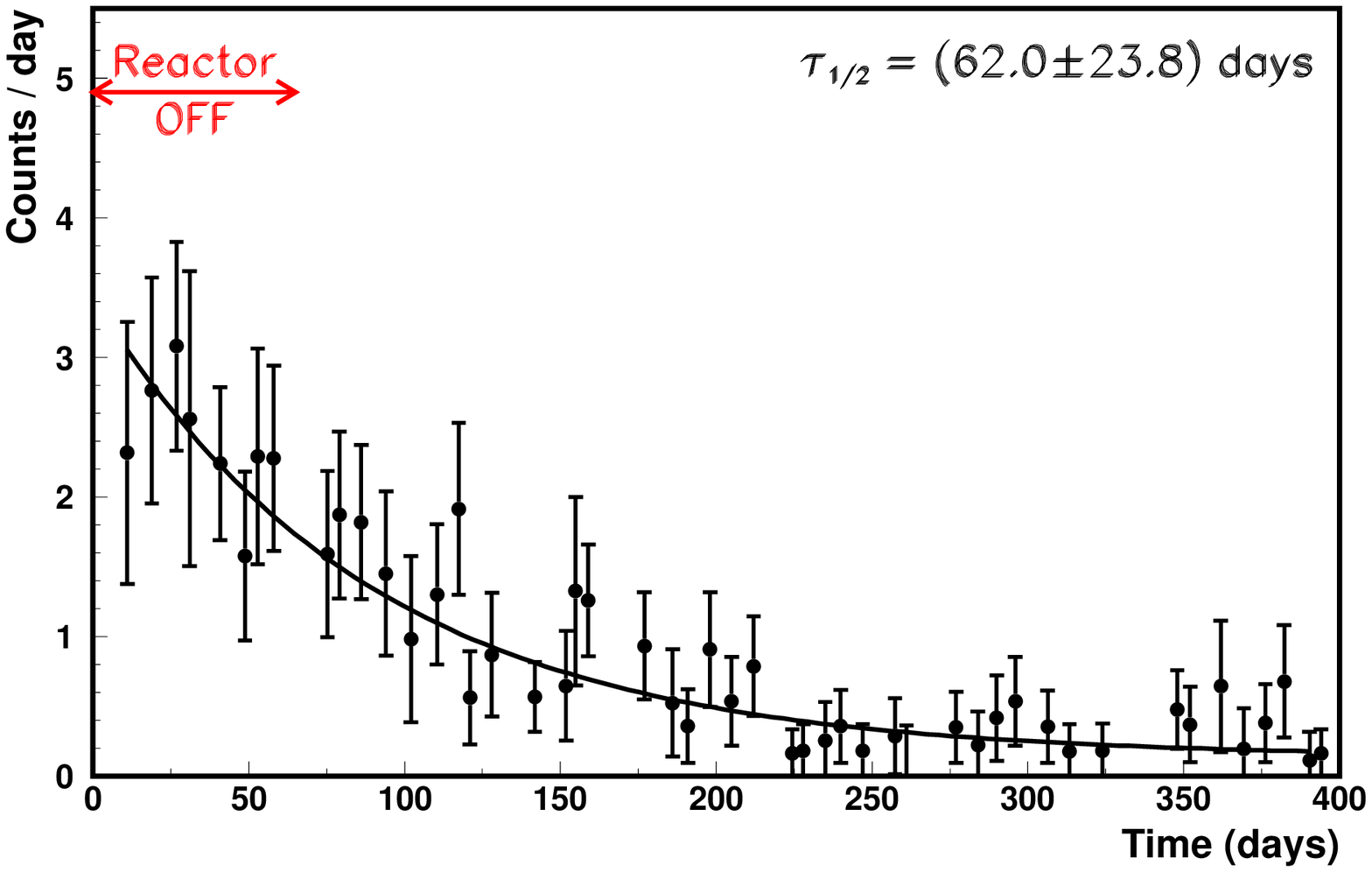}\\
{\bf (b)}\\
\includegraphics[width=10cm]{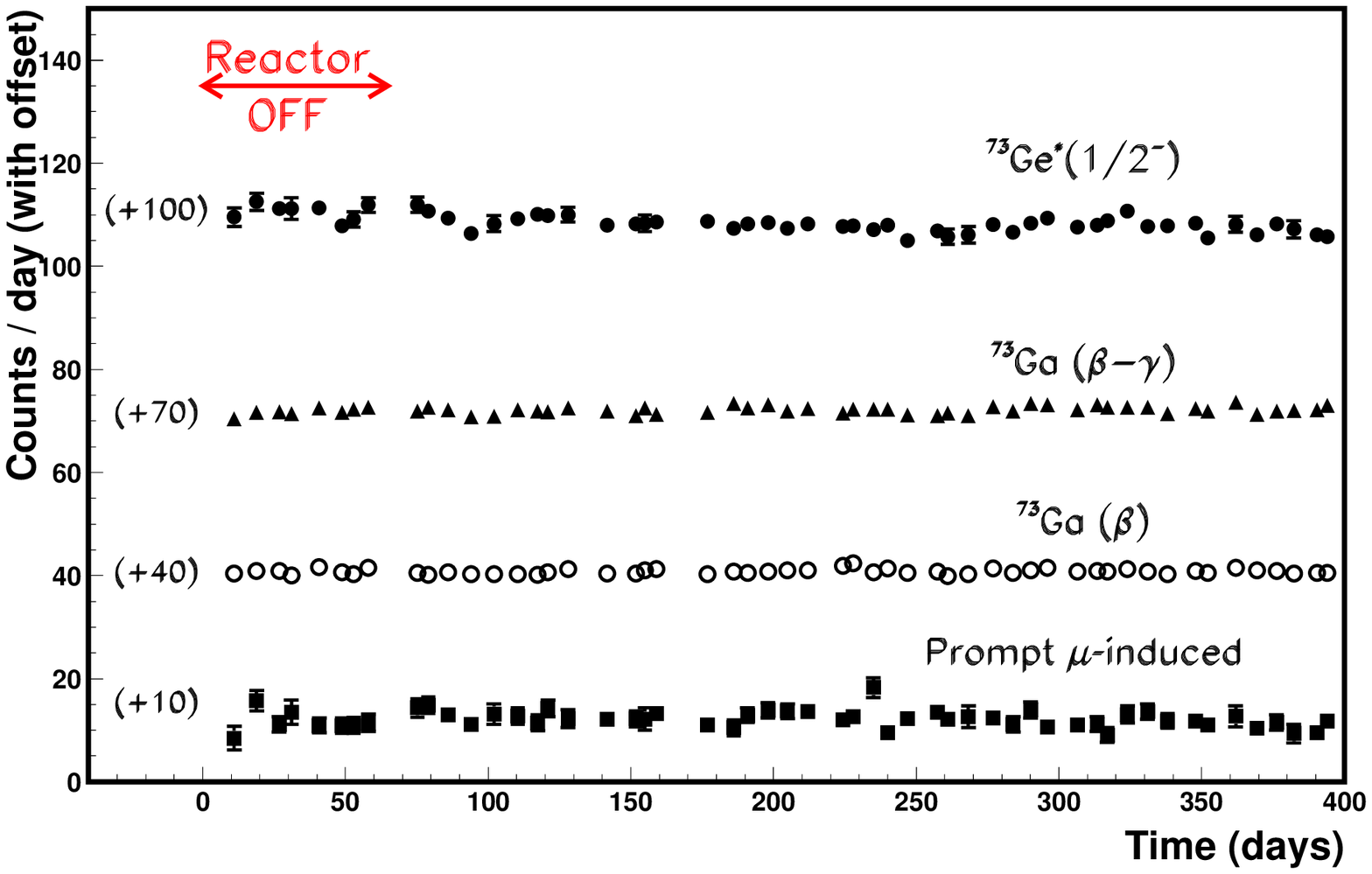}
\end{center}
\caption{
Time plot for P3 data taking for event rates from
(a) Ge X-rays, showing that 
they are due to electron capture
of $^{73}$As to $\ge73m$; and
(b) other channels whose rates are stable.
Offsets are introduced in (b) for visualization
purposes. The data under ``$\ge73m$'' denotes 
the steady-state production rate of $\ge73m$
after the $^{73}$As decays are taken into account.
}
\label{timeplot}
\end{figure}

\begin{figure}
\begin{center}
{\bf (a)}\\
\includegraphics[width=10cm]{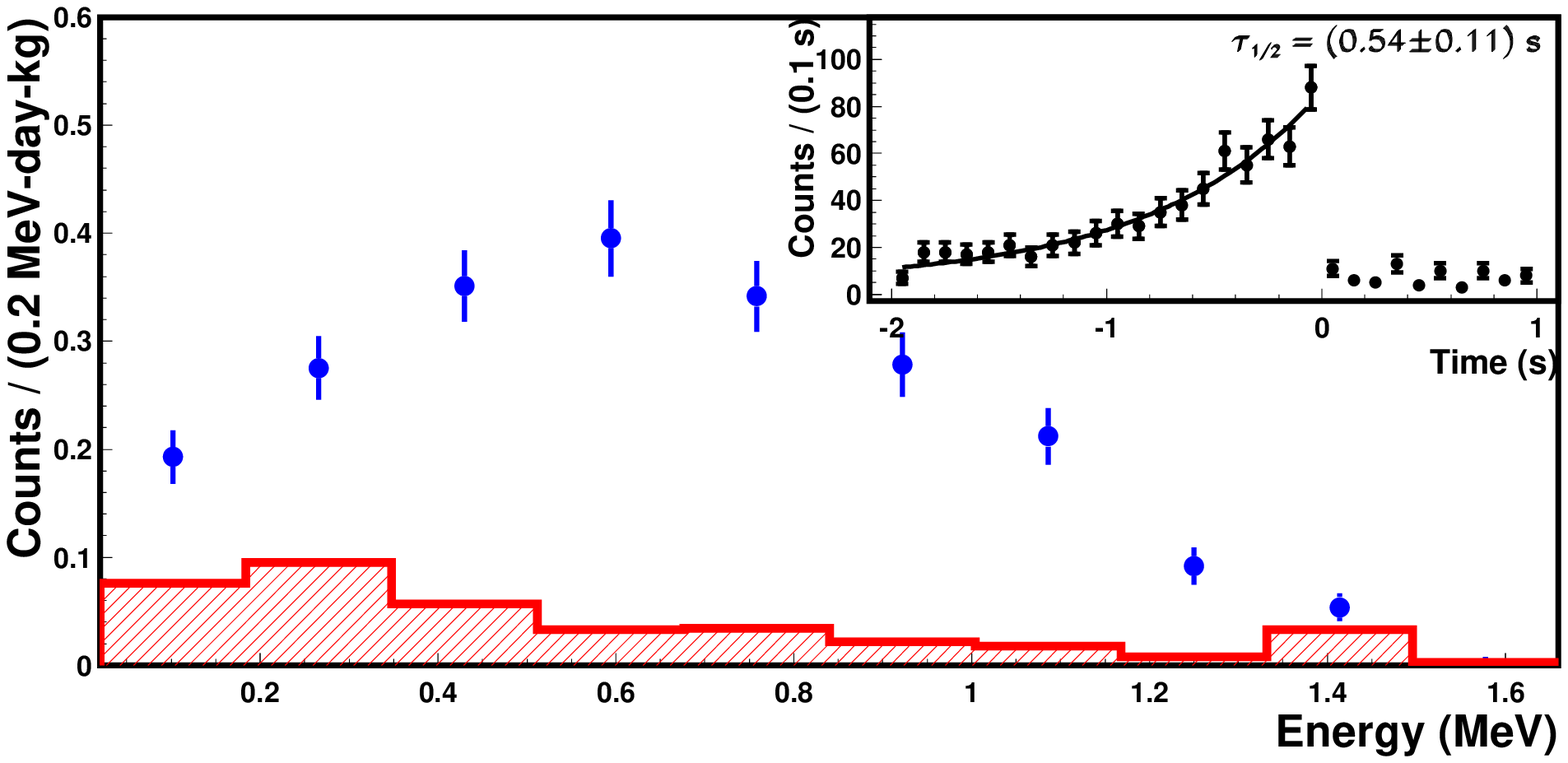}\\
{\bf (b)}\\
\includegraphics[width=10cm]{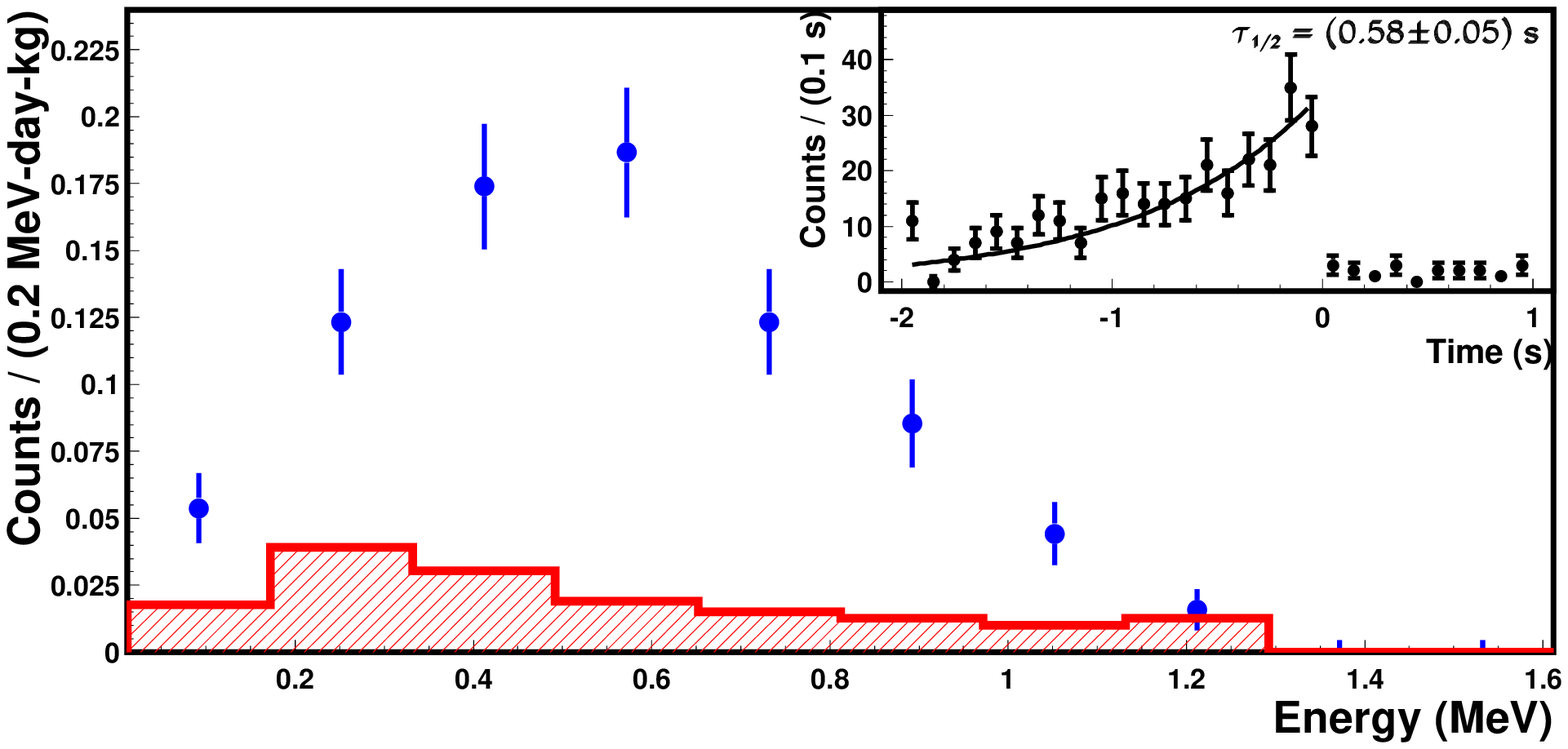}\\
{\bf (c)}\\
\includegraphics[width=10cm]{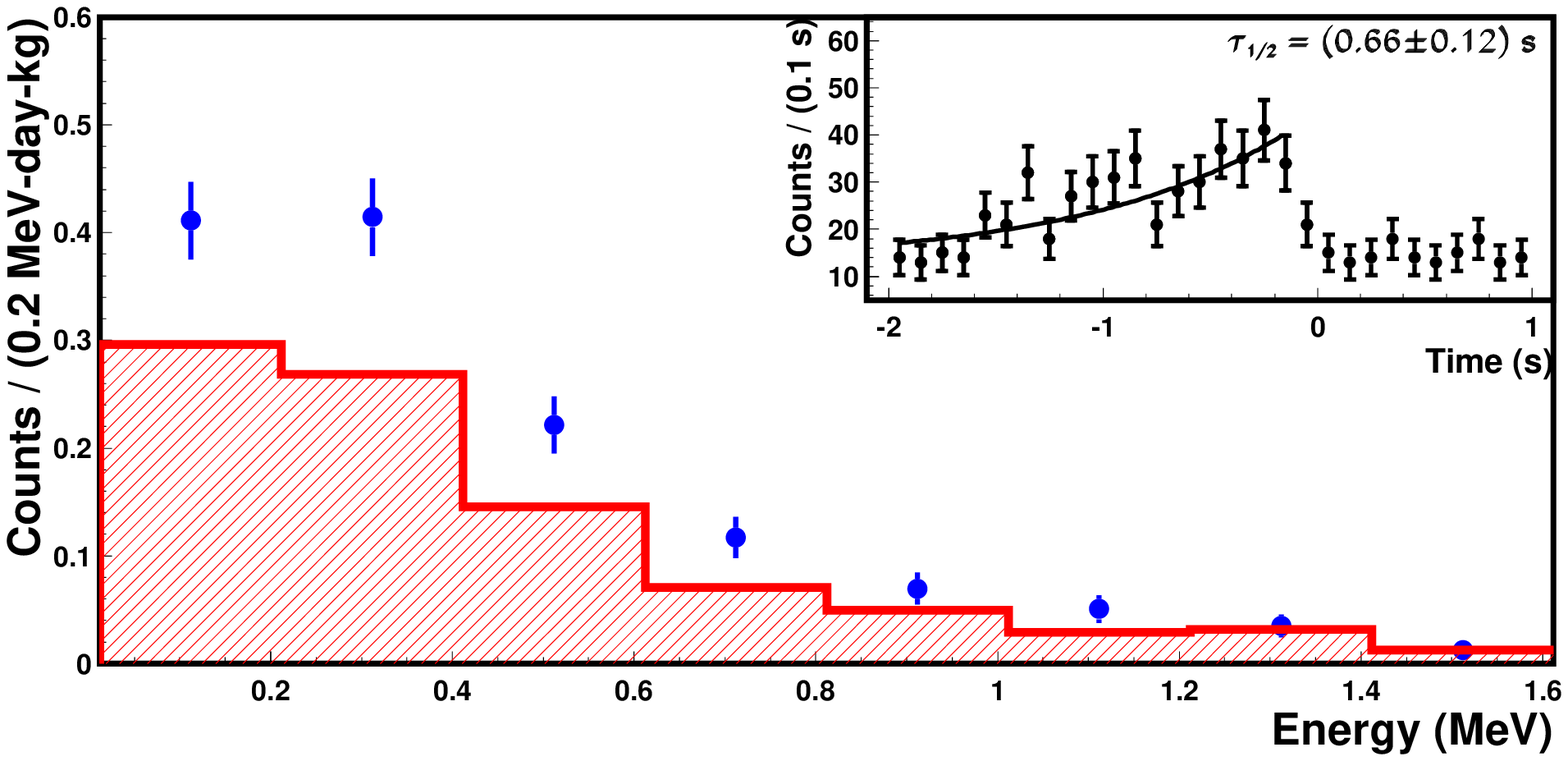}
\end{center}
\caption{
Measured $\phigea$ (red histogram)
and $\phigeb$ (blue data points) 
spectra in the 1~MeV region:
(a) after CRV+ACV cuts;
(b) after CRV cut, with signals at ACV
corresponding to $\rm{E_{\gamma} \sim 300 ~ keV}$;
(c) after CRV cut, with signals at ACV
corresponding to $\rm{E_{\gamma} \neq 300 ~ keV}$.
Inset shows the timing distribution
of the $\phigeb$ events.
}
\label{beta}
\end{figure}

\item {\bf $^{73}$Ga :}\\
The $\phigeb$ and $\phigea$ spectra 
at the 1~MeV region for P3
are depicted 
in Figures~\ref{beta}a-c: 
(a) after CRV+ACV cuts;
(b) after CRV cut, with signals at ACV
corresponding to $\rm{E_{\gamma} \sim 300 ~ keV}$;
and
(c) after CRV cut, with signals at ACV
corresponding to $\rm{E_{\gamma} \neq 300 ~ keV}$.
The timing distribution of the events 
is displayed in the inset of Figures~\ref{beta}a-c,
verifying that $\ge73m$ nuclei were produced.
These events are consistent with 
$\beta^{-}$-emissions from 
$^{73}$Ga in the HPGe:
\begin{equation}
{ \rm ^{73}Ga }  \rightarrow  
{ \rm \ge73m + \gamma 's + e^- + \bar{\nu}_e } 
 ~~ { \rm ( ~ Q = 1.59~MeV ~ ;  ~
\tau _{\frac{1}{2}} = 4.86~h ~ ) }~~ .
\end{equation}

Some of the events in Figure~\ref{beta}a are due to
$\beta ^-$ decays which directly fed the $\ge73m$ state
at a branching ratio (BR) of 7\%.
In addition, the $\beta ^-$ decays can also populate
the various excited states, the dominant level
of which is the $\frac{3}{2}^-$ state at BR=78\%,
followed by $\gamma$-ray emissions
at $\rm{E_{\gamma} = 297 ~ keV}$. 
The subsequent $\gamma$'s were
fully absorbed by the HPGe in about 57\% 
of the decays, 
as indicated in Table~\ref{prodsummary}.
These events also contribute to
Figure~\ref{beta}a.
The HPGe spectra where the
$\rm{E_{\gamma} = 297 ~ keV}$ 
photons
were fully and partially absorbed
by the NaI(Tl) AC detector
are depicted in 
Figures~\ref{beta}b\&c, respectively. 
The $\phinaib$ and $\phinaia$ spectra
corresponding to the samples of Figure~\ref{beta}b
are displayed in
Figure~\ref{naispect}, showing
the detection  of the line at  
$\rm{E_{\gamma} = 297 ~ keV}$.

\begin{figure}[hbt]
\begin{center}
\includegraphics[width=10cm]{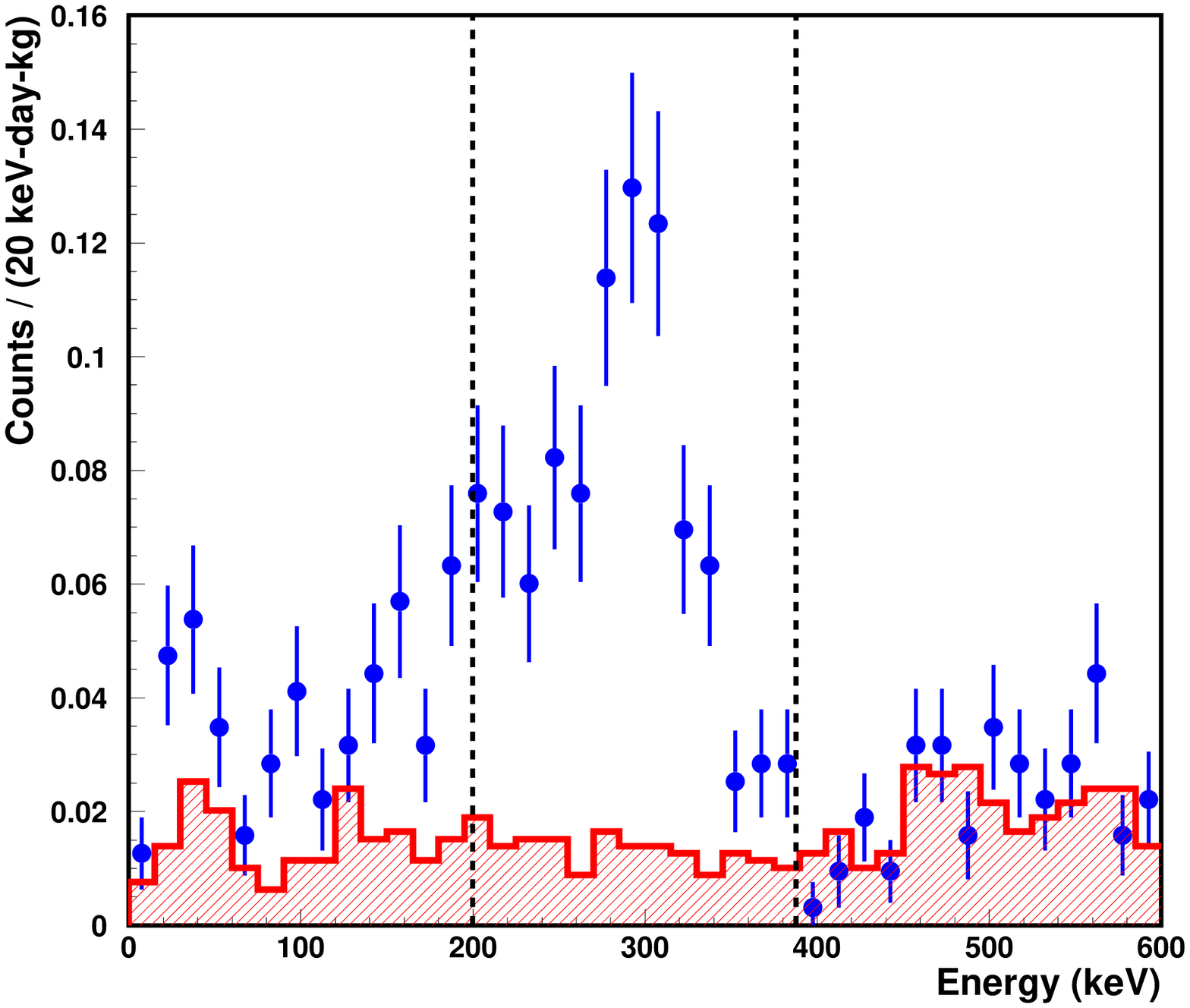}
\end{center}
\caption{
The $\phinaib$ (blue data points)
and $\phinaia$ (red histogram) spectra taken with 
the NaI(Tl) ACV detector.
The photopeak at 300~keV can be tagged by the
specified cuts, giving rise to the
spectra of Figure~\ref{beta}b.
}
\label{naispect}
\end{figure}

\item {\bf Prompt Cosmic-Induced $\ge73m$ :}\\
About 23\% of the production of $\ge73m$ nuclei
were in coincidence with a CRV tag.
These events are usually
characterized by large energy depositions  and
saturate the electronics in
the HPGe or ACV detectors,
as depicted in Figure~\ref{cosmic}a.
For instance, one of the channels is
the neutron capture on $^{72}$Ge:
\begin{equation}
\rm{
n + ^{72}Ge \rightarrow
\ge73m + \gamma 's 
}
\end{equation}
where photons with total energy of
as much as 8~MeV 
are generated.
Only (15.8$\pm$4.0)\% of the events 
were of low energy where the
electronics remained unsaturated
(see also Figures~\ref{cosmic}a\&b).
There was an excess of events at about 300~keV
for $\phinaib$ over $\phinaia$ due
to $\gamma$-ray emissions from the 
$\frac{3}{2} ^-$(364~keV)
excited state of $^{73}$Ge. 
The corresponding energy depositions
in the HPGe were less than 1~MeV.
These events represent evidence of
excitation of  $^{73}$Ge$^*$ through the
interactions of high-energy neutrons produced
by cosmic-ray induced
spallations in the ambient materials. \\
Among the tagged $\ge73m$ events,
1.4\% and (23\%X0.84)$\sim$19.3\% 
are, respectively, those without and with CRV tags
where the total energy depositions
in the HPGe and AC detectors exceed the
end point of $^{73}$Ga $\beta ^-$ decays.
This is consistent with an independent measurement 
of the CRV inefficiency of about 7\%~\cite{texonomagmom},
due to geometrical coverage and hardware inefficiency.
Accordingly, it can be concluded that all 
$\ge73m$ nuclei produced with energy depositions above
1.6~MeV are associated with prompt cosmic rays.

\begin{figure}[hbt]
\begin{center}
{\bf (a)}\\
\includegraphics[width=10cm]{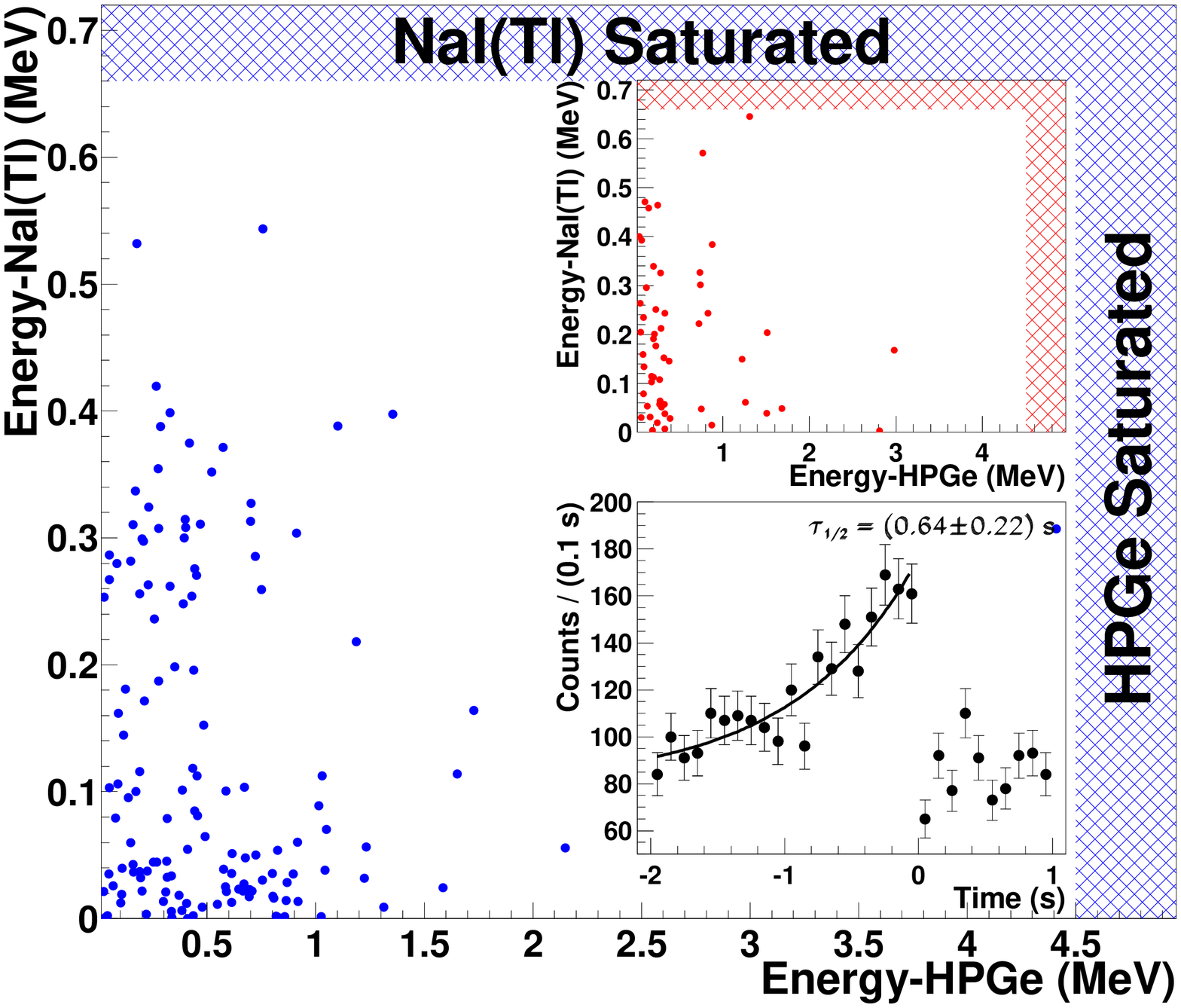}\\
{\bf (b)}\\
\includegraphics[width=10cm]{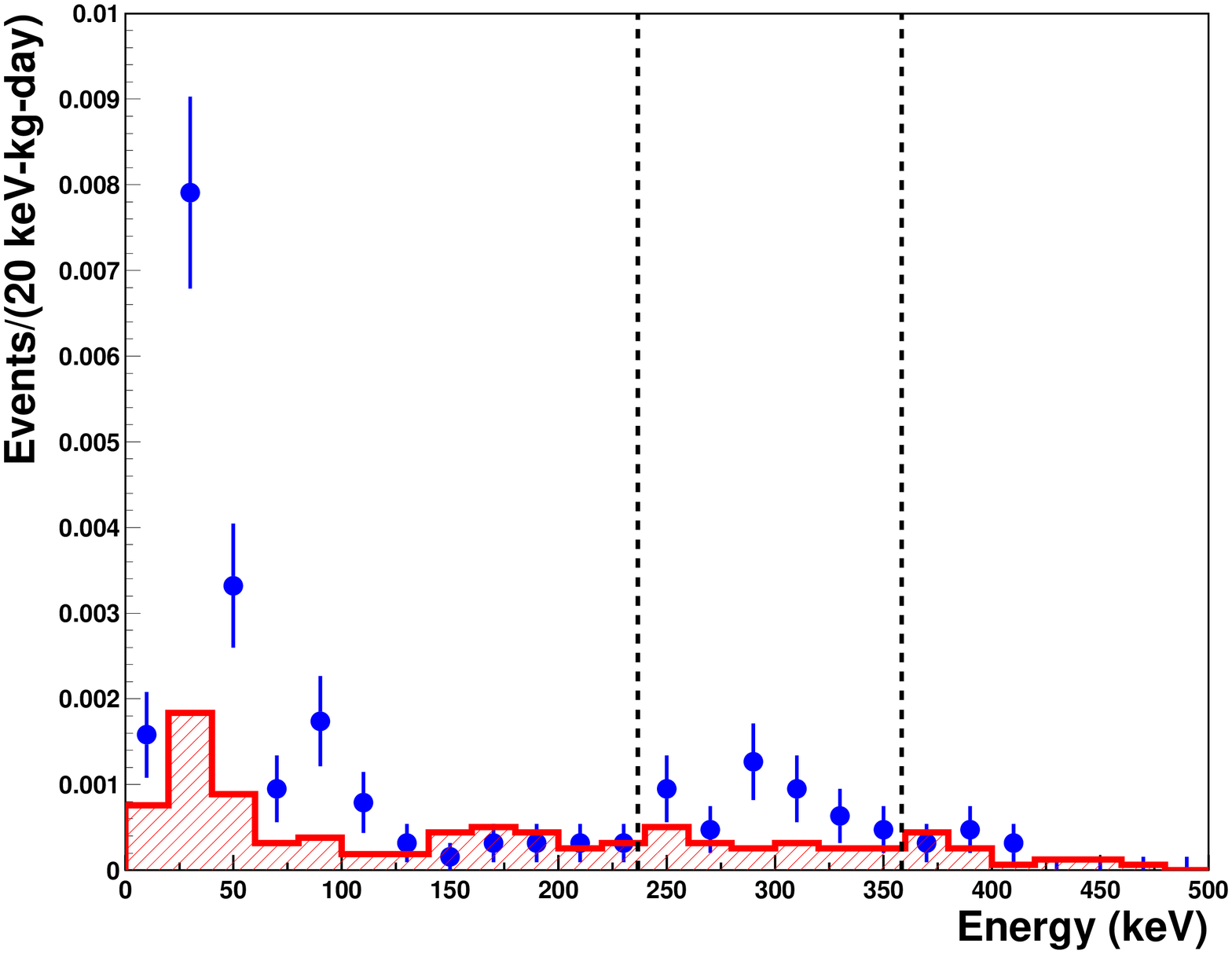}
\end{center}
\caption{
(a)
Scatter plots 
$\phinaib$ versus $\phigeb$ (blue)
and  
$\phinaia$ versus $\phigea$ (red in the inset)
for events with cosmic-ray tag. 
Only events without
saturating the electronics are shown,
while the saturated ones are represented by
the two bands.
The timing distribution of the 
$\phigeb$ events 
is displayed in the inset.
(b)  
The corresponding $\phinaib$ (blue data points)
and $\phinaia$ (red histogram) spectra for
the non-saturated events.
The peak at 300~keV 
is evidence of cosmic-ray induced 
production of $\rm{^{73} Ge ^*}$ excited states.
}
\label{cosmic}
\end{figure}

\end{enumerate}


After corrections on DAQ dead time
and selection efficiencies,
the identified production channel is about 93\% 
of the tagged $\ge73m$ decays. 
The missing events can be attributed to those
which do not satisfy the DAQ trigger condition of
having more than 5~keV energy deposition in the HPGe.
An example of such channels is 
the production of $\ge73m$ or other 
$^{73}$Ge$^*$ excited states via
the excitation of high-energy neutrons,
followed by complete escape of the final-state $\gamma$'s
from the HPGe.

All the identified background is cosmic-ray 
induced, even though the decay time scales are
vastly different: order of 100~days for $^{73}$As,
order of 10~hours for $^{73}$Ga and prompt signatures
for those with a CRV tag.
As illustrated in Figures~\ref{timeplot}a\&b,
the background rates of $^{73}$As show the 
characteristic 80.3~day decay half-life, 
while the rates for the other background
channels are constant with time.
Accordingly, this background is expected to
be greatly suppressed in an underground laboratory
where the cosmic-ray fluxes are attenuated.
The residual background will be due to 
interactions of the surviving cosmic rays
as well as to the fast neutrons produced by
($\alpha$,n) reactions
through natural ambient radioactivity.

\section{Studies of Possible Neutrino-Induced Interactions}
\label{sect::nint}
 
Neutrino interactions on nuclei were studied
in counter experiments
only for a few light
isotopes ($^1$H~\cite{reines}, $^2$H~\cite{deuteron,sno}
and $^{12}$C~\cite{carbon}). 
For heavy nuclei, these were observed
in radiochemical
experiments ($^{37}$Cl~\cite{snucl} 
and $^{71}$Ga~\cite{snuga}) but without timing and
spectral information.
Detailed theoretical work was 
confined mostly to these isotopes.
Establishing more experimentally accessible
detection channels will be of importance in the
studies of nuclear structure and neutrino physics.
For instance, interactions with
lower threshold or better resolution
than the $\nuebar$-p inverse
$\beta ^-$ decay and
$\nu$-d disintegration processes will open
up new windows of investigations.

With reactor $\nuebar$ as probes,
neutrino-induced $\ge73m$ productions
would manifest
themselves as excess of events for the
Reactor ON spectra [$\phigeb (ON)$]
over that of Reactor OFF [$\phigeb (OFF)$].
Studies were performed on the
tagged $\ge73m$ events discussed
in previous sections.


\begin{table}[hbt]
\small{
\begin{tabular}{lccccccc}
\hline
NCEX  & $\Delta$ & E$_{\gamma }$  & Transitions & 
$\rm{f _{\nu}^{\Delta }}$ 
& $\rm{\epsilon _{Ge}}$ & $\rm{\mathcal{R}_{\nu}^{NC}}$ & 
$\rm{ \left\langle \sigma _{\nu } ^{NC} \right\rangle }$  \\
Channel & (MeV)  &  (MeV)  &  &  &  
&  $\rm{( kg^{-1} day^{-1} ) }$   &  ($\rm{cm^{2}}$)   \\ \hline
$\rm{^{73}Ge^* ( 3/2 ^- )}$ & 0.364 & 0.297 
& $\rm{ 9/2^{+} \rightarrow 3/2^{-}}$ 
&  0.82
& 0.49 & $< 2.00 \times  10^{-2}$ & $< 1.13 \times 10^{-43}$  \\ 
$\rm{^{73}Ge^* ( 3/2 ^- )}$ & 0.392 & 0.325 
& $\rm{ 9/2^{+} \rightarrow 3/2^{-}}$ 
&  0.80
& 0.45 & $< 4.36 \times  10^{-2}$ & $< 2.72 \times  10^{-43}$  \\ 
$\rm{^{73}Ge^* (1/2 ^+ )}$ & 0.555 & 0.488 & 
$\rm{ 9/2^{+} \rightarrow 1/2^{+}}$ 
&  0.72
& 0.19 & $< 2.21 \times  10^{-2}$ & $< 3.24 \times  10^{-43}$ \\ 
$\rm{^{73}Ge^* ( 1/2 ^- )}$ & 1.132 & 1.065 & 
$\rm{ 9/2^{+} \rightarrow 1/2^{-}}$ 
&  0.45
& 0.15 & $< 3.12 \times  10^{-2}$ & $<  5.96 \times  10^{-43}$ \\ \hline\hline 
& & & & & &   \\
$\nu$CC & Q &  & Transition & $\rm{f _{\nu}^{Thr}}$ 
& $\epsilon _{Ge}$ & $\rm{\mathcal{R}_{\nu}^{CC}}$ & 
$\rm{\left\langle \sigma _{\nu } ^{CC} \right\rangle} $    \\
Channel &  (MeV)  &  &  &  &  
&  $\rm{ ( kg^{-1} day^{-1} ) }$   &  $\rm{( cm^{2} )}$      \\ \hline
$\rm{^{73}Ga ( 3/2 ^- )}$ & 1.665  &  & 
$\rm{ 9/2^{+} \rightarrow 3/2^{-}}$ 
&  0.13
& 0.68 & $<$~0.43 &  $< 1.78 \times 10^{-42}$ \\ \hline\hline
\end{tabular}
}
\caption{
Summary of the neutrino-induced
NCEX and $\nu$CC studies, showing 90\% CL limits
on event rates and average cross sections. 
}
\label{tab01}
\end{table}


\subsection{Neutral-Current Nuclear Excitation}

Neutrino-induced neutral current processes have been observed
in the disintegration of the deuteron
with reactor $\nuebar$~\cite{deuteron}
and in the SNO experiment for solar $\nu _e$~\cite{sno}.
For heavier isotopes, the interaction
proceeds through neutral current 
excitation (NCEX) on nuclei via 
inelastic scattering
\begin{equation}
\rm{
\nuebar ~+ ~(A,Z)~ 
\rightarrow ~\nuebar ~+~ (A,Z)^*  ~ .
}
\end{equation}
This process 
was observed only in the 
case of $^{12}$C~\cite{carbon}
using accelerator neutrinos at
energy at the order of O(10~MeV).
Excitations at lower energies using 
reactor neutrinos have 
been studied theoretically~\cite{hclee,donnelly} 
but were not experimentally observed.
This has been proposed as a detection
channel for solar neutrinos for the
isotope $^{11}$B~\cite{borex}
and Dark Matter$-$WIMPs~\cite{wimpncex}.
The NCEX processes are sensitive
to the axial isoscalar component of the 
weak neutral currents and the strange 
quark content of the nucleon~\cite{strangequark}.

The average cross section of the NCEX interactions 
in a neutrino beam with spectrum $\phi_{\nu} (E_{\nu})$ 
is given by, using the conventions of Eq.~6 of
Ref.~\cite{hclee}: 
\begin{equation}
\label{eq::csmean}
\rm{
\left\langle \sigma _{\nu}^{NC}\right\rangle=
\frac{\mathcal{\int}_{\Delta}^{\infty }
\sigma _{\nu}^{NC}(E_{\nu })
\phi_{\nu}(E_{\nu })
dE_{\nu }}
{ \Phi_{\nu}^{Total} }
}
\end{equation}
where $\Delta $ is the nuclear excitation energy,
and 
\begin{equation}
\rm{
\Phi_{\nu}^{Total}  = 
\mathcal{\int}_{0}^{\infty }
\phi_{\nu}(E_{\nu })
dE_{\nu } 
}
\end{equation}
is the total neutrino flux.
The energy dependence of 
the interaction cross section varies as~\cite{hclee}
\begin{equation}
\label{reactor_cross}
\rm{
\sigma_{\nu}^{NC} (E _{\nu }) \propto
(E _{\nu }-\Delta )^{2} ~~ ,
}
\end{equation}
where
the proportional constant depends
on the weak couplings and
nuclear matrix elements.
The observed event rate is accordingly
\begin{equation}
\label{eq::rate}
\rm{
\mathcal{R}_{\nu}^{NC}=
\left\langle \sigma _{\nu}^{NC}\right\rangle 
\cdot \Phi_{\nu}^{Total} \cdot N_{Ge} \cdot \epsilon_{Ge}   
}
\end{equation}
where $N_{Ge}$  is the number 
of $^{73}$Ge nuclei
and
$\epsilon_{Ge} $ is 
the  detection efficiency,
which can be evaluated by simulations.

\begin{figure}[hbt]
\begin{center}
\includegraphics[width=12cm]{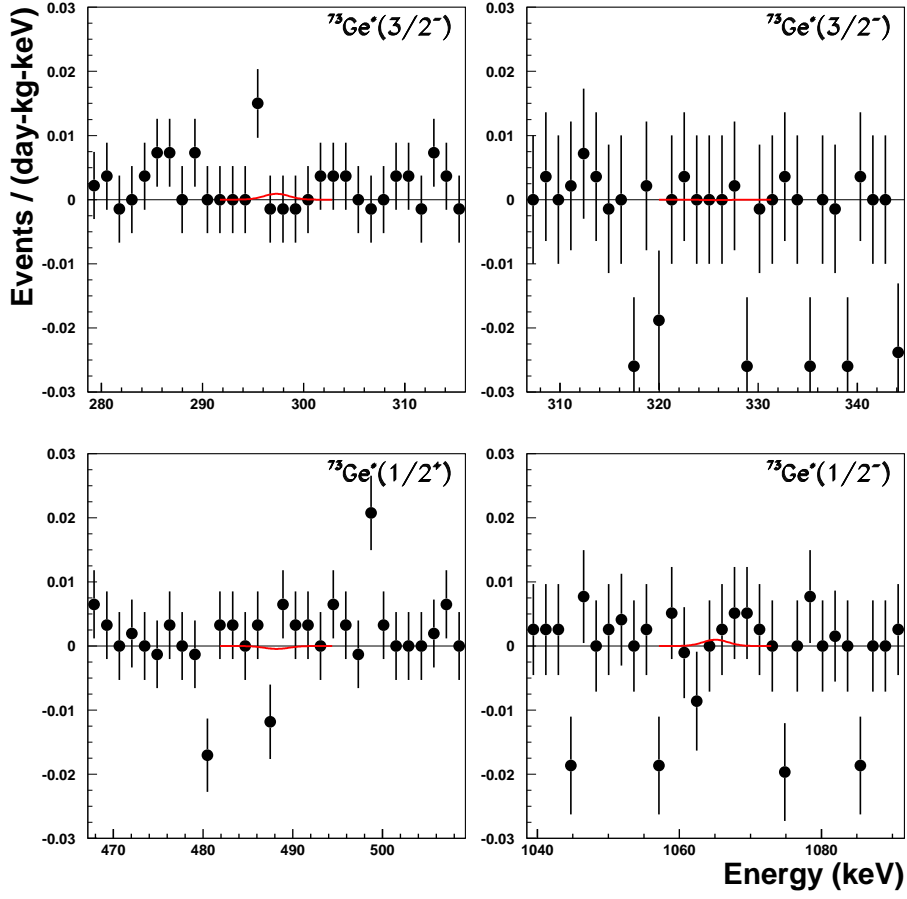}
\end{center}
\caption{
The [$\phigeb(ON) - \phigeb(OFF)$]  residual
spectra for the four 
candidate channels for neutral-current excitation 
with P3 data.
The Gaussian best-fits are
superimposed.
}
\label{ncexresid}
\end{figure}

The experimental signatures of NCEX  
for (A,Z)=$^{73}$Ge
are the presence of mono-energetic 
lines at specified energies in
the Reactor [$\phigeb(ON) - \phigeb(OFF)$] residual spectra.
The candidate $^{73}$Ge-NCEX channels and their respective 
$\epsilon _{Ge}$ are listed in Table~\ref{tab01}.
Also shown is the fraction of the 
reactor neutrino flux
above the kinematics threshold, given by
\begin{equation}
\label{eq::fract}
\rm{
f_{\nu}^{\Delta} = 
\frac{
\mathcal{\int}_{\Delta}^{\infty }
\phi_{\nu}(E_{\nu })
dE_{\nu } }
{ \Phi_{\nu}^{Total} } ~~~ .
}
\end{equation}
The residual spectra of the candidate transitions are
depicted in Figures~\ref{ncexresid}a-d.
No excess of events were observed.
Limits  on $\rm{\mathcal{R} _{\nu} ^{NC}}$, and 
consequently on
$\rm{\left\langle \sigma _{\nu } ^{NC} \right\rangle }$ 
at 90\% CL
were derived and tabulated in Table~\ref{tab01}.
The limiting sensitivities
on $\rm{ \left\langle \sigma _{\nu } ^{NC} \right\rangle }$
are typically factors of $\sim$$10^{2}$ worse than
the typical range 
of $\rm{\sim 10^{-45}cm ~ ^2}$
predicted for the various isotopes~\cite{hclee,donnelly}.
The dominant background are those of Figure~\ref{beta}a,
where all energy of the $\beta$-$\gamma$ emissions 
following $^{73}$Ga decays were deposited in the HPGe.

\subsection{Charged-Current Inverse Beta Decays}

The neutrino-induced inverse $\beta$-decay reaction 
on the proton was the process on which
the first observation of neutrinos was 
based~\cite{reines}.
Subsequently, there were several generations
of oscillation experiments with reactor
neutrinos that relied on this interaction.
In addition, neutrino-induced charged-current processes 
were observed through the disintegration of
the deuteron with reactor $\nuebar$~\cite{deuteron}
and in the SNO experiment
for solar $\nu _e$~\cite{sno}.
For heavy nuclei,
charged-current interactions 
were observed only in solar $\nu_e$  
through radiochemical experiments on 
$^{35}$Cl~\cite{snucl} and $^{71}$Ga~\cite{snuga}.
There are still no successful real-time
counter experiments yet, though there
are intensive R\&D efforts 
towards this goal~\cite{lens}.
The charged-current ($\nu$CC) inverse $\beta$-decay
interaction for $\nuebar$ on heavy nuclei such as $^{73}$Ge
is given by
\begin{equation}
\label{eq::nucc}
\rm{
\nuebar~+ ^{73}Ge \rightarrow ~e^{+}~+ ^{73}Ga^{*}  ~~.
}
\end{equation}
Cases for other heavy isotopes were discussed in 
connection to the detection 
of low energy~$\nuebar$~from the Earth~\cite{nugeo}.
However, there are no experimental 
studies so far for these processes.

Signatures of $\nu$CC in $^{73}$Ge 
manifest themselves as excess of $^{73}$Ga decay events
for the Reactor ON over OFF periods.
The $\nu$CC rate ($\rm{\mathcal{R}_{\nu}^{CC}}$) 
is related to the average cross section via
\begin{equation} 
\rm{
\mathcal{R}_{\nu}^{CC}=
\left\langle \sigma _{\nu}^{CC}\right \rangle 
\cdot \Phi_{\nu}^{Total} 
\cdot N_{Ge} \cdot \epsilon_{Ge}  ~~~.
}
\end{equation}
The interaction cross section varies with neutrino energy as~\cite{nugeo}
\begin{equation}
\rm{
\sigma _{\nu}^{CC} ( E_{\nu} ) \propto 
 F ( Z ,  E_{+} )   \cdot 
E_{+} \cdot
\sqrt { E_{+} ^2 - m_e^2  } 
}
\end{equation}
where $\rm{E_{+} = E_{\nu} - Q - m_e}$ is the positron energy,
and $\rm{F ( Z, E_{+} )}$ is the known nuclear 
Coulomb correction factor, and 
$\rm{Q = 1.67 ~MeV}$ is the Q-value
for the $\nu$CC interaction of Eq.~\ref{eq::nucc}. 

As listed in Table~\ref{tab01},
the residual $^{73}$Ga decay rate for 
the combined P1 and P3 periods is 
$R_{\nu}^{CC}$=${\rm ( - 0.94 \pm 0.67 ) ~ day^{-1} kg^{-1} }$,
from which the 90\% CL limits on $R_{\nu}^{CC}$ and 
$\left\langle \sigma _{\nu}^{CC} \right\rangle$
were derived. 
The fractional $\nuebar$-flux 
($\rm{f ^{Thr} _{\nu}}$) follows the
same definition as Eq.~\ref{eq::fract},
with threshold given by $\rm{ Thr = Q + m_e }$.
Similar to the case for NCEX, the sensitivities
are limited by cosmic-ray induced $^{73}$Ga 
and can therefore be enhanced in an underground location.
There are no calculations on $\nu$CC rates
with reactor neutrinos on heavy nuclei.
Extrapolations from theoretical 
estimates on geo-neutrinos~\cite{nugeo}
suggest a general cross-section range of 
$\rm{ \sim 10^{-44} ~ cm ^2}$
and hence factors of $\sim$$10^{2}$ 
more stringent than the experimental 
bounds.


\section{Summary and Prospects}

We made a thorough study on the decay signatures 
of the $\ge73m$ metastable state in a well-shielded reactor
laboratory at a shallow depth of about 30~mwe. 
An unambiguous event-by-event tag of such decays was 
demonstrated, and studies of the signals within
two seconds before
the tag provide information on their production channels.

Searches for possible neutrino-induced nuclear transitions
give rise to sensitivity limits
which are typically
factors of $\sim 10^2$ worse than
the general predicted range. The background channels are
all cosmic-ray induced, the most relevant one being
the decays of $^{73}$Ga. Consequently, the sensitivities
can be greatly enhanced in an underground location. 
Physics experiments have been conducted in 
the past at the
underground laboratory at the 
Krasnoyarsk reactor~\cite{krasnoyarsk}, 
though this facility is no longer available. 
Additional boost in the capabilities
can be provided by position-sensitive segmented HPGe
which can distinguish single-site from multi-site events.

Ton-scale isotopically-pure Ge-based detectors 
have been proposed and considered in 
forthcoming double-beta decay experiments~\cite{dbdge}
in the case of $^{76}$Ge,
as well as cold dark-matter searches~\cite{cdmge}
where $^{73}$Ge is ideal
for studying the spin-dependent interactions.
Background control and suppression
are crucial to the success of these projects.
Sophisticated procedures have been developed
in the course of the R\&D work.
The unique three-fold timing correlations of the
$\ge73m$ system demonstrated in this article
will further enhance the background rejection
capabilities.

A one-ton  isotopically-pure
$^{73}$Ge detector located
15~m from a 3~GW reactor core
would record $\sim$16 
neutrino-induced NCEX events per day 
at the typical predicted cross-section range of 
${\rm 10^{-45} ~ cm^2 }$~\cite{hclee,donnelly}.
The neutron flux in an underground site 
can be attenuated by a typical
factor of $10^{3}$ or more~\cite{ugdneutron}.
Therefore, a similar
suppression in the $^{73}$Ga background
can be expected.
The background levels of $\rm{ < 0.01 ~ kg^{-1} day^{-1}}$
from Table~\ref{tab01} for natural HPGe at surface
would therefore imply a range of 
$\rm{ < 0.1 ~ ton^{-1} day^{-1} }$ 
for pure $^{73}$Ge detector underground.
The detection of NCEX would be realistic
and possible $-$ if an underground power reactor
would be available.

Measurement of the
low-energy solar neutrinos
at a threshold low enough ($\rm{< 423 ~ keV}$)
to include the pp branch 
with the NCEX processes  
has important complementarity to the on-going efforts
towards detection of the charged-current interactions~\cite{lens}.
Using simple scaling between reactor and solar
neutrino fluxes and spectra,
the typical predicted cross-section 
ranges of Refs.~\cite{hclee,donnelly}
correspond to a solar-$\nu$ induced NCEX rate of 
$\sim 16 ~ \rm{ton^{-1} year^{-1} }$
in a $^{73}$Ge detector. 
This is comparable to the expected
background range in an underground location.
Further optimizations on background control
and suppression 
would make such a process observable.

\section{Acknowledgments}

This work is supported by
contracts 93-2112-M-001-030,
94-2112-M-001-028 and
95-2119-M-001-028
from the National Science Council, Taiwan.

\end{document}